%% file: TWEPP_2019_Proceedings.tex
\documentclass{PoS}

\usepackage{textcomp}
\usepackage{lineno}
\usepackage{floatrow}
\usepackage{siunitx}
\usepackage{graphicx}
\usepackage{setspace}
\usepackage{color,soul}
\usepackage[belowskip=-10pt,aboveskip=5pt]{caption,subcaption}

\title{CLICTD: A monolithic HR-CMOS sensor chip for the CLIC silicon tracker}

\ShortTitle{CLICTD: A monolithic HR-CMOS sensor chip for the CLIC silicon tracker}

\author{\speaker{I. Kremastiotis~}$~^{a,b}${, R. Ballabriga}$~^{a}${, K. Dort}$~^{a,c}${, N. Egidos}$~^{a,d}${, M. Munker}$~^{a}$}
\author{{on behalf of the CLICdp collaboration}\\ \\
	    $~^{a}$CERN, Geneva, Switzerland\\
	    $~^{b}$Karlsruhe Institute of Technology, Karlsruhe, Germany\\
	    $~^{c}$Justus-Liebig-Universitaet Giessen, Giessen, Germany\\
		$~^{d}$University of Barcelona, Barcelona, Spain\\
        E-mail: \email{iraklis.kremastiotis@cern.ch}}

\abstract{The CLIC Tracker Detector (CLICTD) is a monolithic pixelated sensor chip produced in a $180$~nm imaging CMOS process built on a high-resistivity epitaxial layer.
The chip, designed in the context of the CLIC tracking detector study, comprises a matrix of ${16\times128}$ elongated pixels, each measuring ${300\times30}$~\SI{}{\micro\meter\squared}.
To ensure prompt charge collection, every elongated pixel is segmented in eight sub-pixels, each containing a collection diode and a separate analog front-end.
A simultaneous $8$-bit time measurement with $10$~ns time bins and $5$-bit energy measurement with programmable range is performed in the on-pixel digital logic.
The main design aspects as well as the first results from laboratory measurements with the CLICTD chip are presented.}

\FullConference{Topical Workshop on Electronics for Particle Physics\\
		2 - 6 September 2019\\
		Santiago de Compostela, Spain}

\begin{document}
	
\input{./TWEPP_2019_main_text.tex}

\end{document}

%% file: TWEPP_2019_main_text.tex
\newcommand{\latex}{\LaTeX\xspace}
%\lstset{defaultdialect=[LaTeX]TeX}
 
\section{Introduction}
\label{sec:intro}
\vspace{-3mm}

A novel monolithic pixel sensor chip, the CLIC Tracker Detector (CLICTD), is presented. 
The CLICTD chip features a matrix of $128$ rows and $16$ columns with pixels of $300\times30$~\SI{}{\micro\meter\squared}.
The chip was designed according to the requirements for the silicon tracker at the future Compact Linear Collider (CLIC)~\cite{trackerReqs}.
%These requirements \hl{include a time} resolution of $\sim5$~ns, being addressed by an $8$-bit Time of Arrival (ToA) measurement with $10$~ns time bins, and a $5$-bit Time over Threshold (ToT) measurement for time walk correction. 
These requirements include a Time of Arrival (ToA) measurement with $10$~ns time bins, and a $5$-bit Time over Threshold (ToT) measurement for time walk correction.
Moreover, the $5$-bit ToT measurement (which has a programmable range from $0.6$~to~$4.8$~\SI{}{\micro\second}) enables a precise hit position interpolation that is needed to reach a single point resolution of \SI{7}{\micro\meter} along the transverse plane with the pitch of \SI{30}{\micro\meter}.
Other requirements include a total material budget of $1 - 1.5\%$~X$_{0}$ per detection layer (allowing for $\sim200$~\SI{}{\micro\meter} for the silicon layers) and an average power consumption below \SI{150}{\milli\W}$/$\SI{}{\centi\meter\squared}. 
In order to minimise the average analog power consumption, the analog front-end can be set to a standby power mode in the $20$~ms gaps between subsequent colliding bunch trains (power pulsing), taking advantage of the low duty cycle of the CLIC beam~\cite{clic}. 
The digital power consumption is minimised by means of clock gating. 
%The resulting average power consumption over the CLIC cycle is \SI{5}{\milli\W}$/$\SI{}{\centi\meter\squared} for the matrix, plus $70$~mW for the periphery (for $3\%$ occupancy).
Details on the CLICTD chip design have been presented in~\cite{twepp2018}.

\vspace{-5mm}
\section{Process description}
\label{sec:process}
\vspace{-3mm}

\begin{figure}[htbp]
	\centering
	\begin{subfigure}[b]{.4\textwidth}
		\centering % \begin{center}/\end{center} takes some additional vertical space
		\includegraphics[width=1\textwidth]{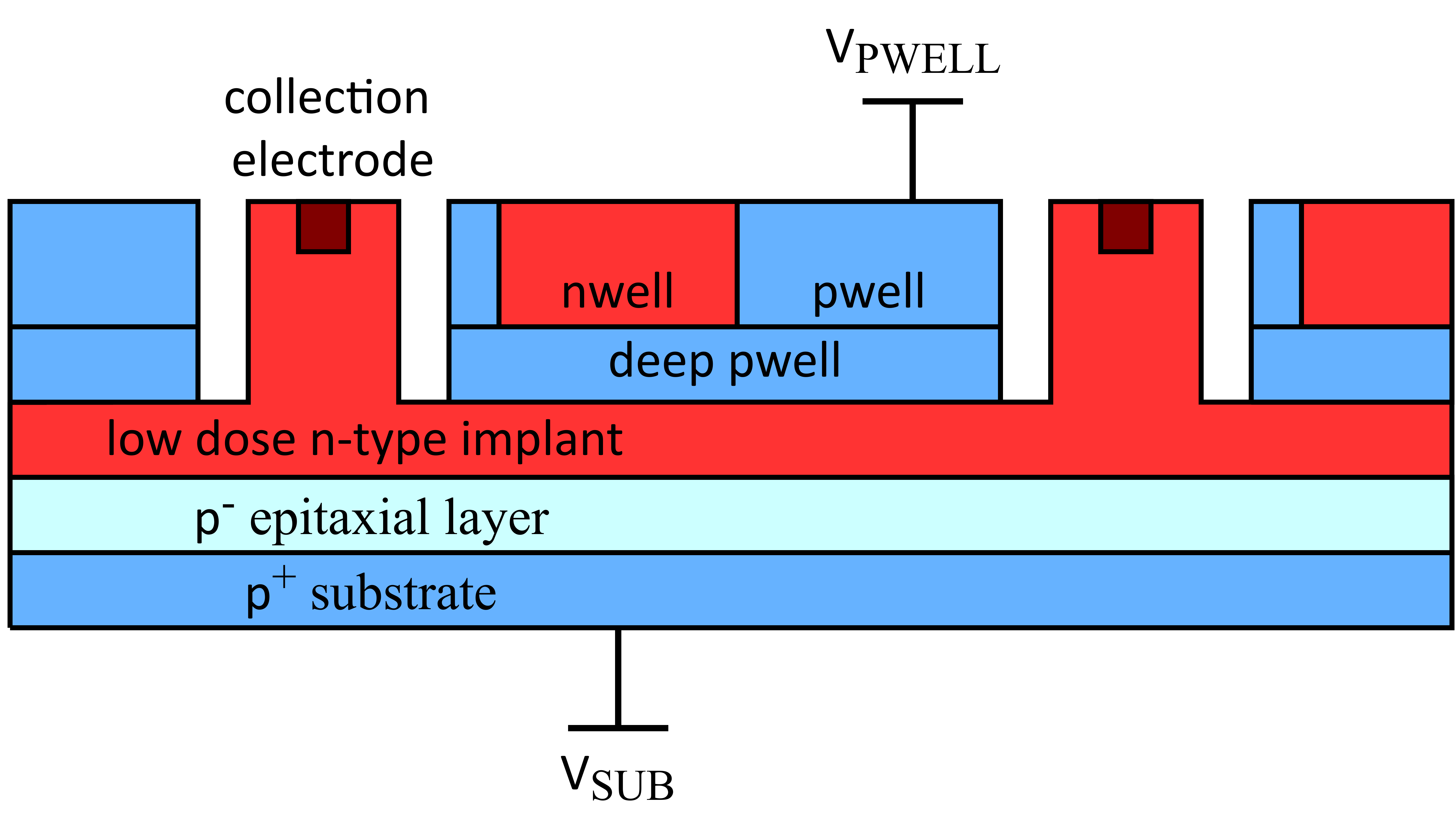}
		\caption{\label{fig:processA}}
	\end{subfigure}
	\begin{subfigure}[b]{.4\textwidth}
		\centering % \begin{center}/\end{center} takes some additional vertical space	
		\includegraphics[width=1\textwidth]{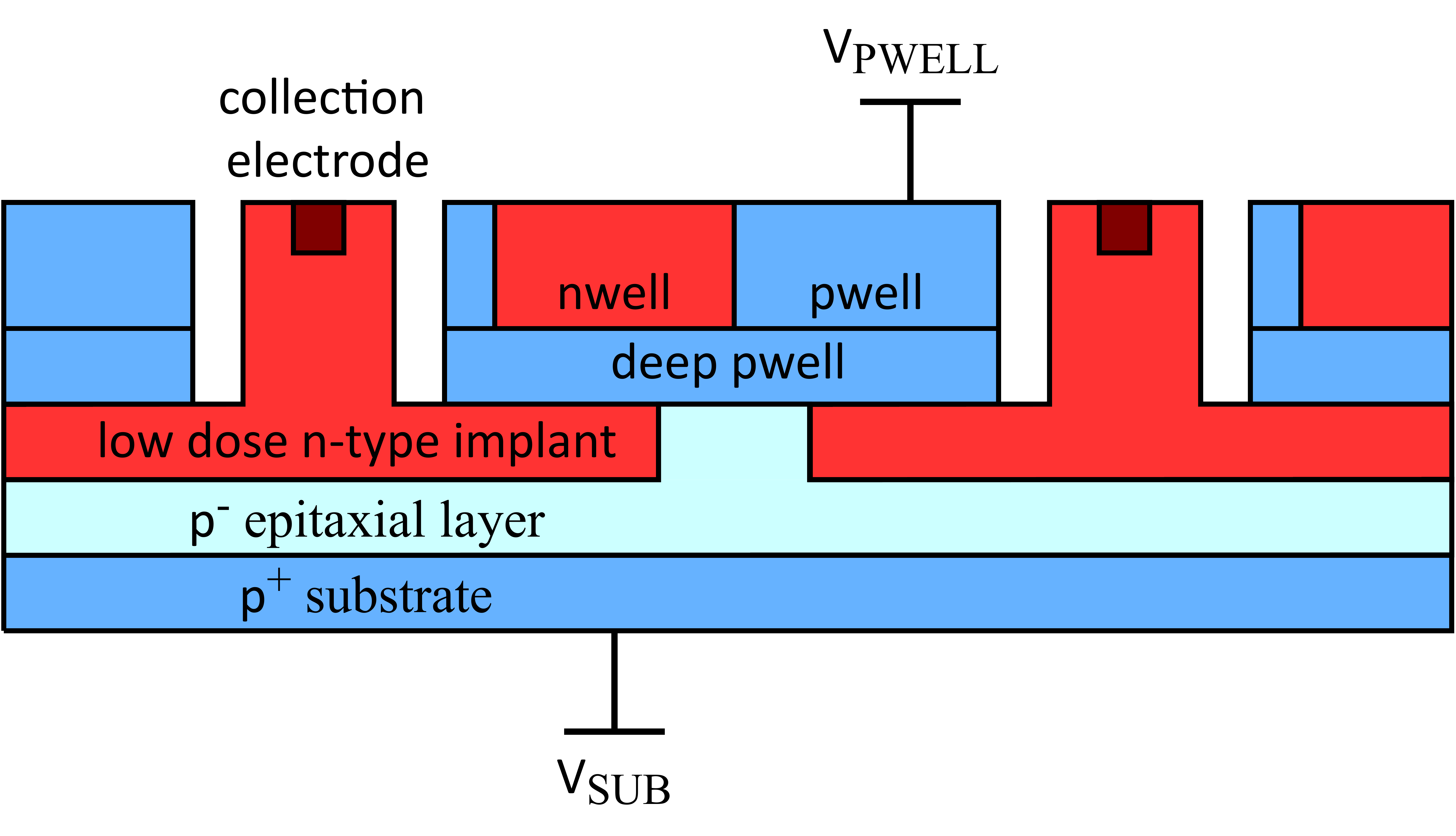}
		\caption{\label{fig:processB}}
	\end{subfigure}
	\caption{\label{fig:processSplit} The two process splits: (\subref{fig:processA}) continuous N-implant, and~(\subref{fig:processB}) gap in N-implant (not to scale).}
\end{figure}

The design was implemented in a $180$~nm High-Resistivity (HR) CMOS imaging process, where a deep P-well is used in order to shield the on-pixel electronics from the collection electrode. 
The signal is collected with a small area N-well on top of a P-type high resistivity epitaxial layer. 
This results in a small capacitance (a few fF) for the collection electrode and helps to minimise the analog power consumption and the noise in the front-end.
Full depletion of the epitaxial layer is achieved by including an additional deep N-type implant.
Two different pixel layouts were implemented and manufactured through a process split on different wafers.
As illustrated in Figure~\ref{fig:processSplit}, the first pixel layout is with a continuous deep N-type implant, and the second with a segmented deep N-type implant to increase the lateral field and thereby reduce the charge collection time~\cite{Munker2019B}.
In addition, reduced charge sharing is expected because of the gap in the N-implant.
The gap is introduced only along the long dimension of the pixel, as the short pixel dimension corresponds to the transverse detector plane, where charge sharing can be beneficial for improving the position resolution through hit position interpolation.
%other dimension the charge sharing is needed for position reconstruction.
%A schematic cross-section of the two process splits is presented in Figure~\ref{fig:processSplit}.

\vspace{-6mm}
\section{The CLICTD prototype chip}
\label{sec:chip}
\vspace{-3mm}

The CLICTD matrix comprises $16\times128$ elongated pixels of $300\times30$~\SI{}{\micro\meter\squared}. 
Each pixel is segmented in eight sub-pixels, each with its individual collection diode and analog front-end, to ensure prompt charge collection.
The diodes are therefore spaced by \SI{37.5}{\micro\meter} along the long pixel direction. 
Every front-end includes level shifter, voltage amplifier, discriminator and a $3$-bit local threshold tuning DAC. 
The level shifter acts as a unity gain buffer and is placed as close to the collection electrode as possible to minimise the input capacitance.
%Simulations show a minimum detectable charge of $93$~$e^{-}$, an in-time charge of $720$~$e^{-}$ (where the time walk remains below $10$~ns) for \SI{210}{\milli\W}$/$\SI{}{\centi\meter\squared} (continuous operation, without power pulsing). 
One bit indicating whether the front-end detected a particle is stored for each sub-pixel, while the simultaneous $8$-bit ToA and $5$-bit ToT measurement is performed in the on-pixel digital logic for the combined output (by means of an "OR" gate) of all eight discriminator outputs.
%\hl{The ToT measurement has a programmable range from $0.6$~to~$4.8$~\SI{}{\micro\second}.}
Thus, the time-stamp of the first hit and the integral ToT of all hits within the same frame are read out for each pixel.
%The slow control is based on the I$^2$C protocol, while a serial readout at $40$~MHz, with a zero suppression algorithm, is employed. 
A serial readout at $40$~MHz is employed.
To reduce the amount of data shifted out of the chip, a compression algorithm is applied. 
The data acquired during the frame are shifted out only for pixels that detected a particle. For pixels that did not detect a particle, only one bit is shifted out.
%The chip was verified using the Universal Verification Methodology (UVM)~\cite{uvm}.

\vspace{-4mm}
\section{Measurement results}
\label{sec:measurements}
\vspace{-3mm}

Once fabricated, the CLICTD chip was characterised using the CaRIBOu readout system~\cite{caribou}.

%Once the CLICTD chip was fabricated, samples were wire-bonded on a custom designed PCB and characterised using the CaRIBOu readout system~\cite{caribou}.
%Figure~\ref{fig:clictd} presents a CLICTD sample, wire-bonded on the PCB and Figure~\ref{fig:caribou} presents the CaRIBOu readout system, including an evaluation kit based on the SoC architecture, an interface board and the PCB where the CLICTD sample is mounted.

%\begin{figure}[htbp]
%	\centering
%	\begin{subfigure}[b]{.4\textwidth}
%		\centering % \begin{center}/\end{center} takes some additional vertical space
%		\includegraphics[width=1\textwidth]{figures/clictdPhoto.png}
%		\caption{\label{fig:clictd}}
%	\end{subfigure}
%	\begin{subfigure}[b]{.4\textwidth}
%		\centering % \begin{center}/\end{center} takes some additional vertical space	
%		\includegraphics[width=1\textwidth]{figures/clictdReadout.jpg}
%		\caption{\label{fig:caribou}}
%	\end{subfigure}
%	\caption{\label{fig:setup} (\subref{fig:clictd}) Photo of the CLICTD chip. (\subref{fig:caribou}) Photo of the CaRIBOu data acquisition system.} 
%\end{figure}

\vspace{-4mm}
\subsection{I-V characteristic}
\label{sec:clictdIV}
\vspace{-2mm}

\begin{figure}[t!]
	\centering
	\begin{subfigure}[b]{.49\textwidth}
		\centering % \begin{center}/\end{center} takes some additional vertical space
		\includegraphics[width=1\textwidth]{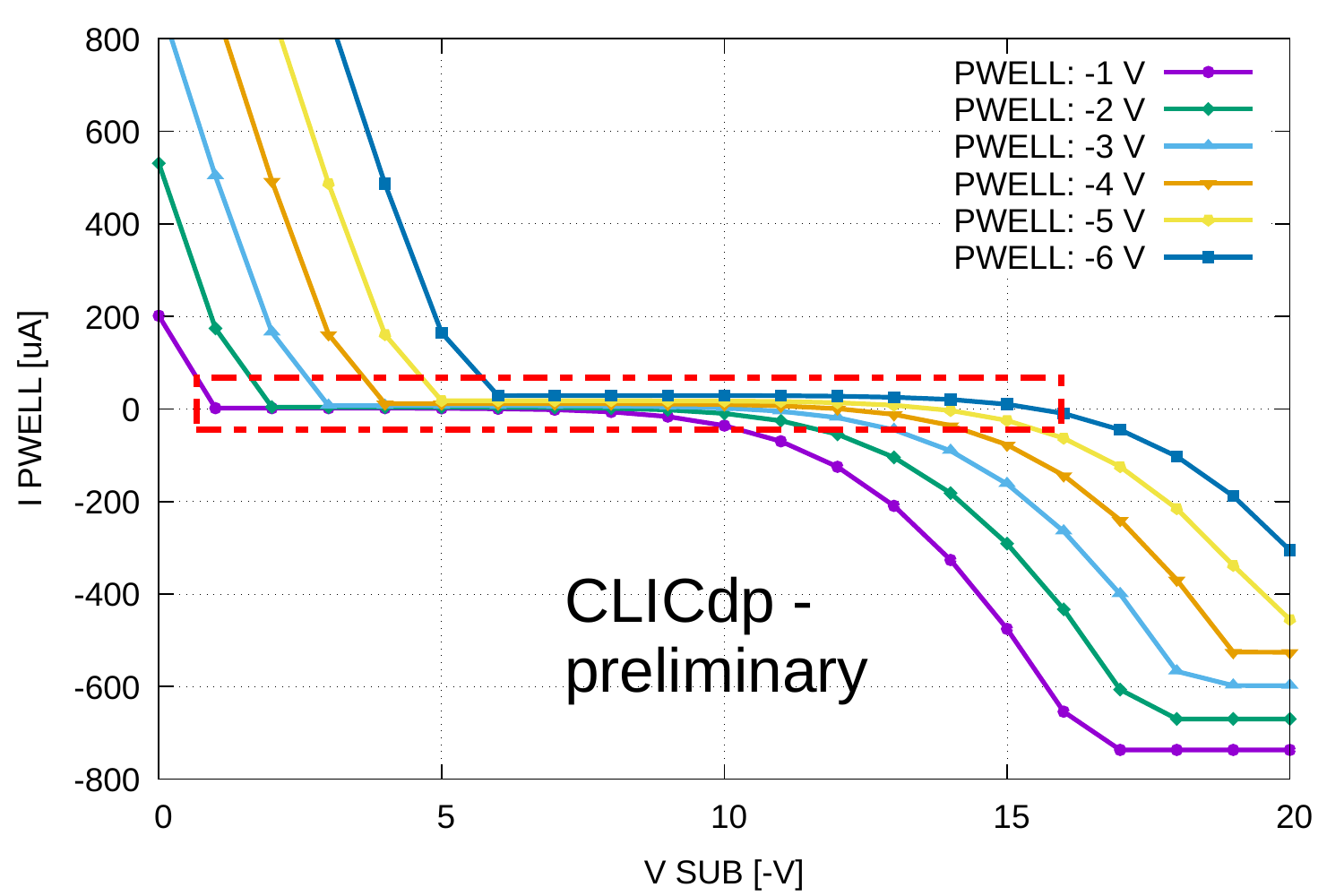}
		\caption{\label{fig:pwellA}}
	\end{subfigure}
	\begin{subfigure}[b]{.49\textwidth}
		\centering % \begin{center}/\end{center} takes some additional vertical space	
		\includegraphics[width=1\textwidth]{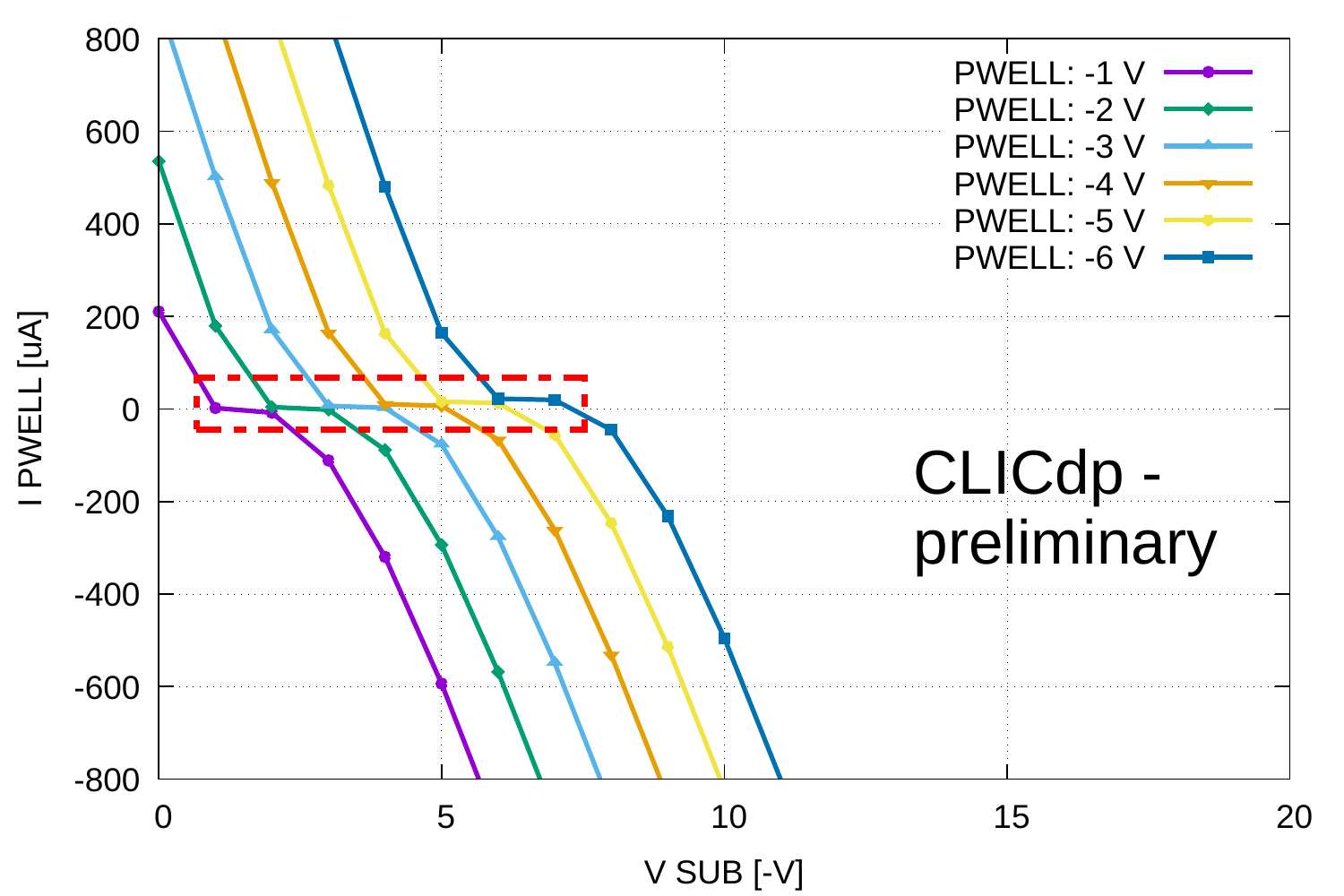}
		\caption{\label{fig:pwellB}}
	\end{subfigure}
	\caption{\label{fig:IV} Leakage current measured at the P-wells in the CLICTD matrix as a function of the substrate bias, for different values of the P-well bias. CLICTD sample from~(\subref{fig:pwellA}) the first process option - continuous N-implant, and~(\subref{fig:pwellB}) the second process option - gap in N-implant.} 
\end{figure}

The sensor bias operation range was determined by measuring its I-V characteristics. 
%The leakage current at the P-wells in the pixel matrix was measured, while scanning the substrate bias voltage (V\textsubscript{SUB} in Figure~\ref{fig:processSplit}), and for fixed values of the P-well bias (V\textsubscript{PWELL} in Figure~\ref{fig:processSplit}) ranging from $-1$ to $-6$~V.
The leakage current at the P-wells in the pixel area (biased at V\textsubscript{PWELL} in Figure~\ref{fig:processSplit}) was measured, while scanning the substrate bias voltage, and for given values of the P-well bias ranging from $-1$ to $-6$~V.
%As illustrated in the plots in Figure~\ref{fig:IV}, depending on the two bias voltages, the I-V characteristic is divided in three different regions. 
The region where the leakage current remains stable and relatively low ($<20$~\SI{}{\micro\ampere} for the full chip) is the region where the sensor can be operated (marked in red in Figure~\ref{fig:IV}). 
For the first pixel layout in Figure~\ref{fig:processA}, the continuous deep N-type implant offers a better isolation between the P-wells and the substrate. 
Therefore, the substrate can be biased to a higher (in absolute value) voltage, as shown in Figure~\ref{fig:pwellA}. 
Due to the reduced isolation originating from the gap in the deep N-type implant in the second pixel layout, the sensor can be operated only when the two nodes are biased at a similar voltage level (Figure~\ref{fig:pwellB}), a result which is in agreement with simulations~\cite{Munker2019B}. 

\vspace{-4mm}
\subsection{Noise measurements and equalisation}
\vspace{-2mm}

\begin{figure}[t!]
	\centering
	\begin{subfigure}[b]{.49\textwidth}
		\centering % \begin{center}/\end{center} takes some additional vertical space
		\includegraphics[width=1\textwidth]{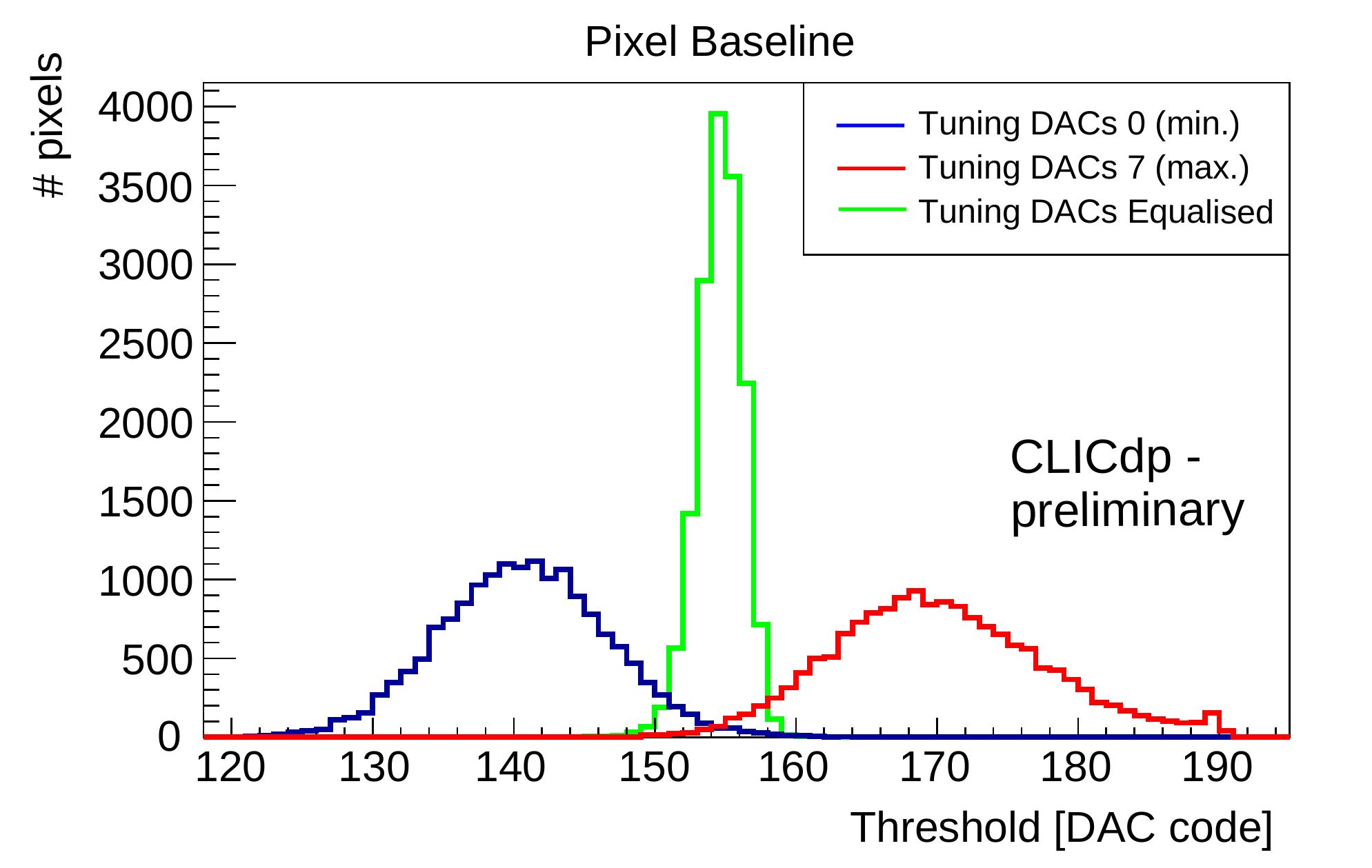}
		\caption{\label{fig:equalisation}}
	\end{subfigure}
	\begin{subfigure}[b]{.42\textwidth}
		\centering % \begin{center}/\end{center} takes some additional vertical space	
		\includegraphics[width=1\textwidth]{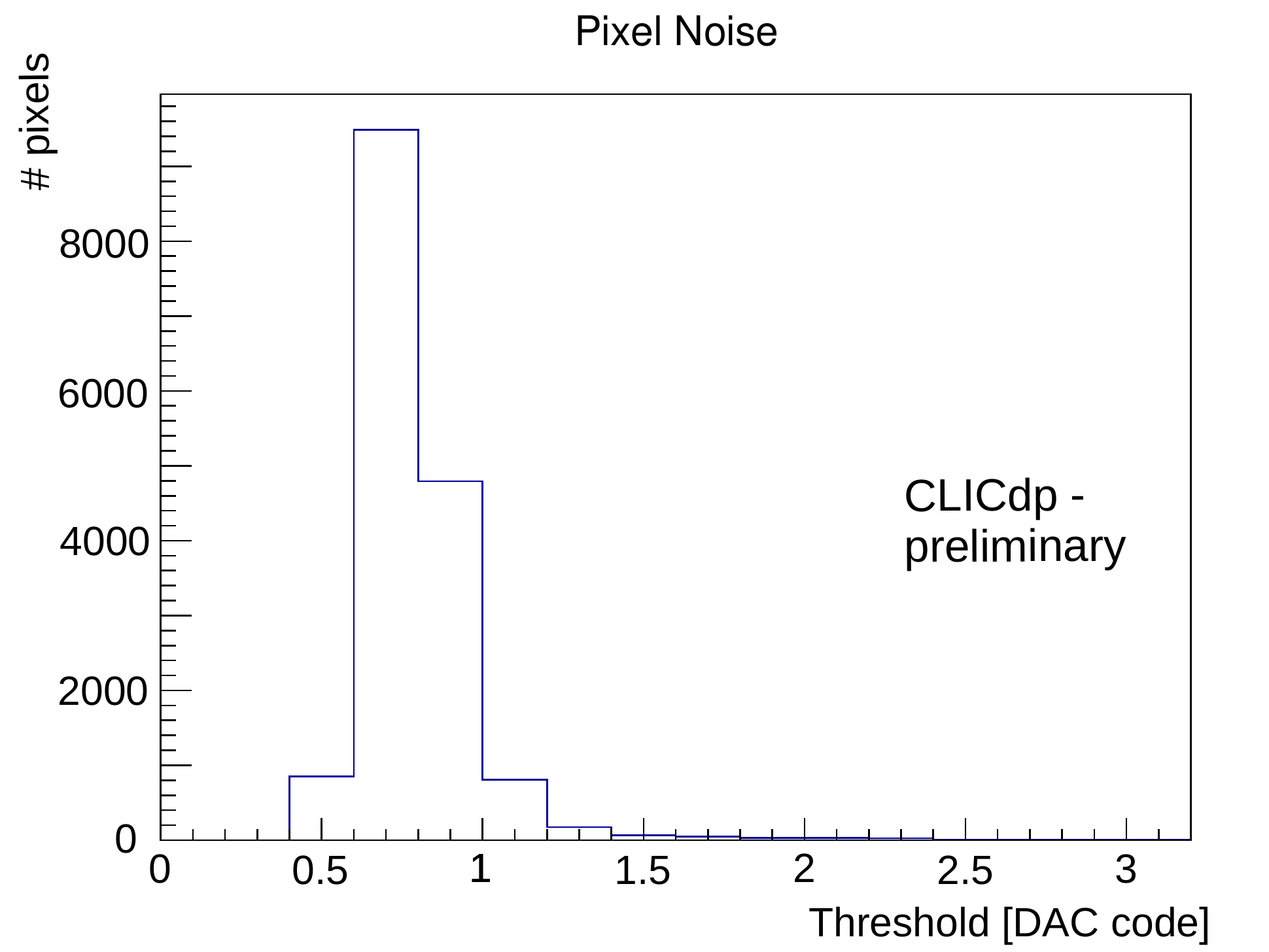}
		\caption{\label{fig:noise}}
	\end{subfigure}
	\caption{\label{fig:eq} (\subref{fig:equalisation}) Distribution of pixel baselines obtained from threshold scans before (blue, red) and after (green) equalisation. (\subref{fig:noise}) Histogram of noise RMS per sub-pixel, in threshold DAC codes.} 
\end{figure}

%As introduced in Section~\ref{sec:chip}, a $3$-bit DAC is placed in each front-end in order to correct for threshold dispersion along the pixel matrix.
During the threshold equalisation for the CLICTD pixel matrix, the baseline for each sub-pixel is extracted as the mean value of the curve produced after scanning the threshold and counting the number of noise hits for each threshold.
%by performing a threshold scan through the noise. 
In order to equalise the matrix on a sub-pixel basis, the scan was performed for each of the eight sub-pixels in the pixel, while the rest of the sub-pixels were masked.
The threshold scans were performed for the lowest and highest code of the $3$-bit threshold tuning DAC in the front-end.
A linear interpolation was applied in order to select the DAC codes for which the threshold dispersion over the matrix is minimised.
Figure~\ref{fig:equalisation} presents the threshold scan for lowest (blue curve), highest (red curve) and equalised (green curve) DAC code.
A histogram of the noise RMS per sub-pixel, extracted in threshold DAC codes as the RMS value of the curve resulting from the threshold scan for each sub-pixel, is plotted in Figure~\ref{fig:noise}.
Based on the threshold DAC step to electrons conversion extracted in Section~\ref{sec:sources}, the threshold dispersion for the equalised CLICTD matrix is about $25$~e$^-$ and the mean value of the noise RMS is $13$~e$^-$.
The above results are from a CLICTD sample with gap in the n-layer, with both the P-wells and the substrate biased to $-2$~V.
%As extracted from fluorescence measurements with iron and copper targets (Section~\ref{sec:sources}), one threshold LSB corresponds to $\sim16$~$e^{-}$.

\vspace{-4mm}
\subsection{Measurements with radiation sources}
\label{sec:sources}
\vspace{-2mm}

\begin{figure}[htbp]
	\centering
	\begin{subfigure}[b]{.45\textwidth}
		\centering % \begin{center}/\end{center} takes some additional vertical space
		\includegraphics[width=1\textwidth]{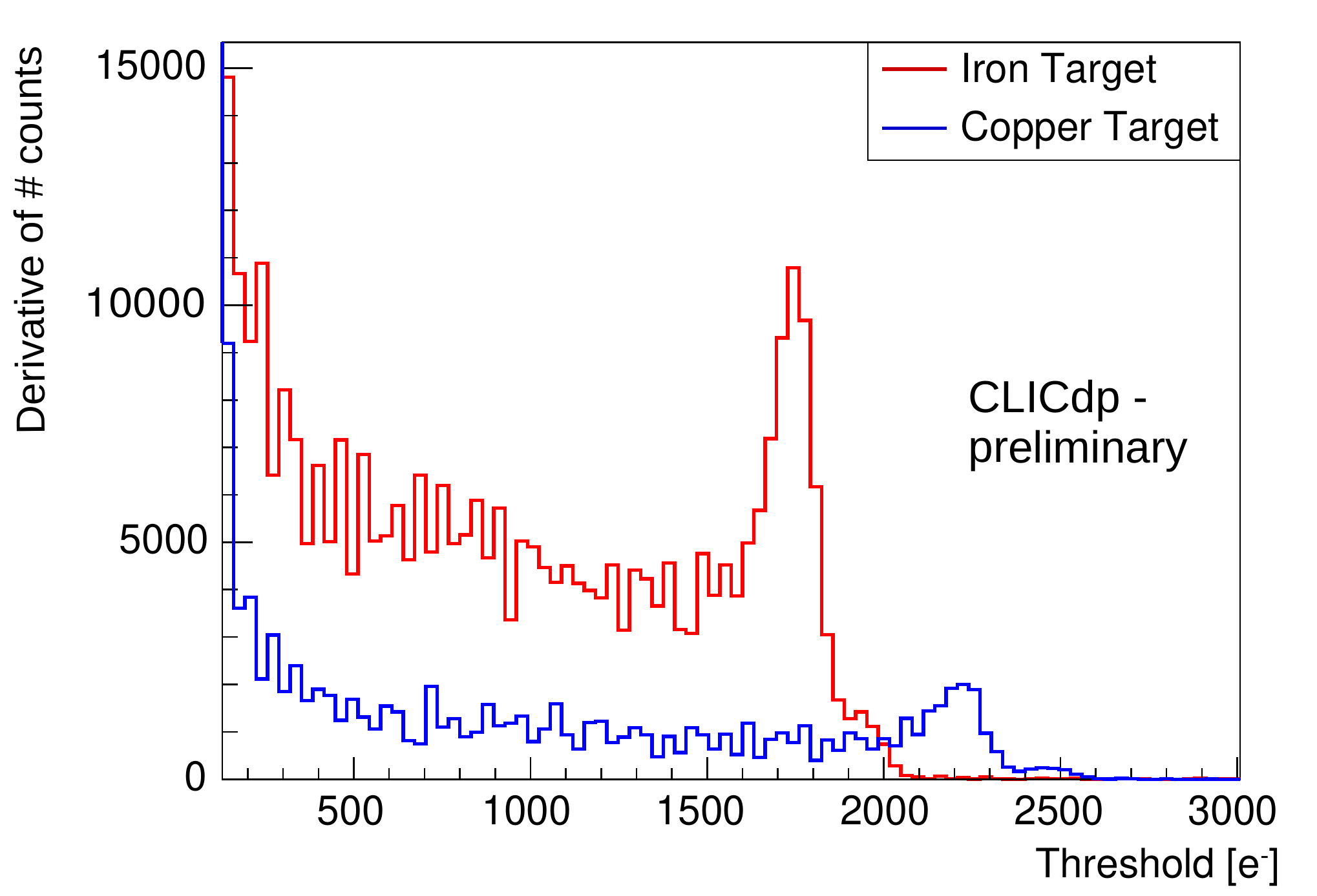}
		\caption{\label{fig:spectra}}
	\end{subfigure}
	\begin{subfigure}[b]{.45\textwidth}
		\centering % \begin{center}/\end{center} takes some additional vertical space	
		\includegraphics[width=1\textwidth]{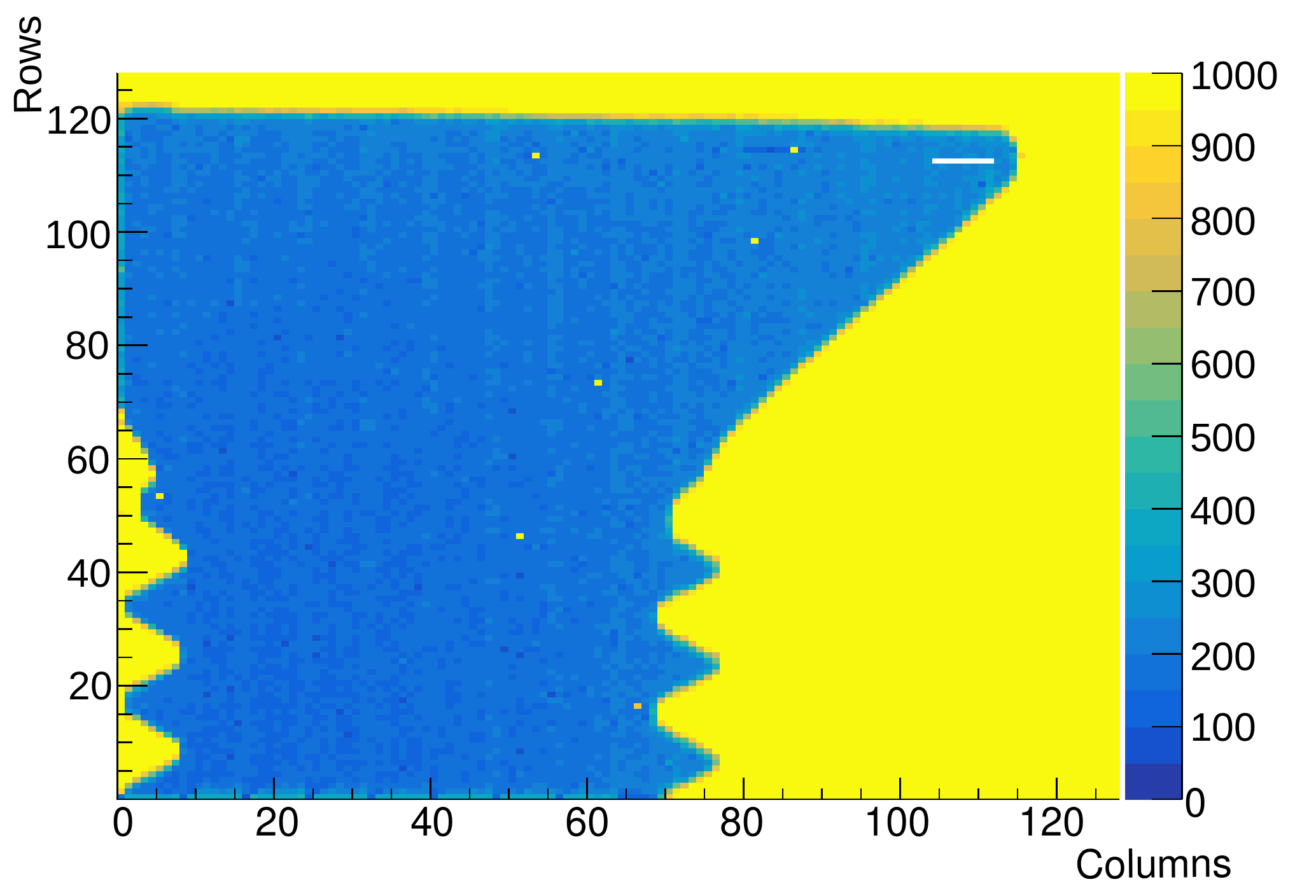}
		\caption{\label{fig:xRay}}
	\end{subfigure}
	\caption{\label{fig:xrays} (\subref{fig:spectra}) Fluorescence spectra using iron (red) and copper (blue) targets, measured with the CLICTD chip. (\subref{fig:xRay}) X-ray image of a screw, recorded with the CLICTD chip.} 
\end{figure}

Figure~\ref{fig:spectra} presents the fluorescence spectra from an iron (red curve) and a copper (blue curve) target, measured with the CLICTD chip using an X-ray beam. 
The CLICTD threshold was scanned during this measurement and the number of detected photons was stored in each pixel and read out for every threshold. 
%The spectrum is obtained from the derivative of the produced occupancy curve.
The produced occupancy curve gives the number of counts above threshold, for each applied threshold step.
For a given beam intensity and frame duration, this number depends on the fluorescence emission spectrum of the target. 
A differentiation of the occupancy curve gives the measured energy spectrum.

From the mean values of the Gaussian fits for the K\textsubscript{$\alpha$} peaks for iron ($6.4$~keV) and copper ($8.04$~keV), and with an electron-hole pair creation energy in silicon of $3.62$~eV, the LSB of the threshold DAC has been extracted to correspond to $\sim16$~e$^{-}$.
The minimum detectable charge for the chip settings applied for this measurement is estimated to be $\sim300$~e$^{-}$.

Using the same setup, an X-ray image of a small item (screw) was taken with the CLICTD chip. 
%For this measurement, a small item (screw) was placed between the X-ray tube and the CLICTD chip. 
The image presented in Figure~\ref{fig:xRay} was produced with $1000$ acquisitions of \SI{4}{\milli\second} shutter length, with equalised matrix.
The number of frames when a hit was detected is plotted for each sub-pixel.
% to obtain an image with a sub-pixel granularity of $37.5\times30$~\SI{}{\micro\meter\squared}.

For the above measurements, a CLICTD sample with continuous N-implant was used, while the reverse bias was set to $-2$~V at the P-wells and to $-4$~V at the substrate.
%As expected from finite element simulations, 
The sensor capacitance (and consequently the front-end performance) strongly depends on the sensor bias~\cite{MunkerThesis}.
The front-end was biased at its default values after chip reset and the operation parameters were not yet tuned for signal-to-noise performance.
Under these bias conditions, the analog part of the matrix consumes $\sim$\SI{170}{\milli\W}$/$\SI{}{\centi\meter\squared}, while the on-pixel digital logic consumes another \SI{240}{\milli\W}$/$\SI{}{\centi\meter\squared}.
%As introduced in Section~\ref{sec:intro}, a power pulsing scheme is applied in order to minimise the average power consumption over the cycle of the CLIC accelerator.
The expected average power consumption of the CLICTD matrix over the cycle of the CLIC accelerator, as extrapolated from static power measurements and simulated values, is $\sim$\SI{5}{\milli\W}$/$\SI{}{\centi\meter\squared}, after applying the power pulsing scheme introduced in Section~\ref{sec:intro}.
In addition, \SI{43}{\milli\W} are consumed by the analog and digital periphery electronics. 
%As the periphery electronics are not power pulsed, this contribution is constant, independent of the applied power pulsing.

\vspace{-4mm}
\section{Summary}
\vspace{-3mm}

A monolithic sensor chip, the CLIC Tracker Detector (CLICTD), designed in the framework of the CLIC silicon tracker study, has been presented.
The chip was produced in a modified CMOS imaging process and features a novel architecture with segmented pixels.
%The chip features a matrix of $16\times128$ elongated pixels of $300\times30$~\SI{}{\micro\meter\squared}. 
%Each pixel can perform a simultaneous $8$-bit ToA and $5$-bit ToT measurement, while it is segmented in eight collection diodes, each with its dedicated front-end electronics, in order to ensure prompt charge collection.
Preliminary results show a threshold dispersion of $25$~e$^-$ and a mean noise RMS of $13$~e$^-$ over the CLICTD matrix, after threshold equalisation.
The average power consumption over the CLIC beam cycle is estimated to be $\sim$\SI{5}{\milli\W}$/$\SI{}{\centi\meter\squared} for the pixel matrix, plus \SI{43}{\milli\W} for the periphery electronics. 

%The first results with the chip have been obtained and further studies, including characterisation in beam tests, are in progress.
%The CLICTD chip is targeting at the CLIC silicon tracker requirements, but it can also serve as a technology demonstrator in order %to explore the different applications where this technology and architecture can be used.

\vspace{-4mm}
\section*{Acknowledgements}
\vspace{-3mm}

This project has received funding from the European Union's Horizon 2020 research and innovation programme under grant agreement No 654168.

%% file: TWEPP_2019_Proceedings.bbl
\vspace{-4mm}
\begin{thebibliography}{99}

\begin{spacing}{0.9}

\bibitem{trackerReqs}
D. Dannheim, A. Nurnberg, \emph{Requirements for the CLIC tracker readout},
\href{https://cds.cern.ch/record/2261066/files/CLICdp-Note-2017-002.pdf}{CLICdp-Note-2017-002}. 2017.

\vspace{-0.5mm}

\bibitem{clic}
\emph{Physics and Detectors at CLIC: CLIC Conceptual Design Report}, \\
edited by L. Linssen, A. Miyamoto, M. Stanitzki, H. Weerts, 
\href{https://cds.cern.ch/record/1425915/files/CERN-2012-003.pdf}{CERN-2012-003}. 2012.

\vspace{-0.5mm}

\bibitem{twepp2018}
I. Kremastiotis et al., \emph{Design of a monolithic HR-CMOS sensor chip for the CLIC silicon tracker},
\href{https://cds.cern.ch/record/2643766/files/CLICdp-Conf-2018-008.pdf}{CLICdp-Conf-2018-008}. 2018.

\vspace{-0.5mm}

%\bibitem{Snoeys2017}
%W. Snoeys et al., \emph{A process modification for CMOS monolithic active pixel sensors for enhanced depletion, timing performance and radiation tolerance},
%\href{https://doi.org/10.1016/j.nima.2017.07.046}{Nucl. Instrum. Meth. A 871 (2017) 90-96}.

\bibitem{Munker2019B}
M. Munker et al., \emph{Simulations of CMOS pixel sensors with a small collection electrode, improved for a faster charge collection and increased radiation tolerance},
\href{https://iopscience.iop.org/article/10.1088/1748-0221/14/05/C05013}{JINST 14  C05013}. 2019.

\vspace{-0.5mm}

%\bibitem{uvm}
%A. Fiergolski, \emph{Simulation Environment Based on the Universal Verification Methodology}, \\
%\href{http://iopscience.iop.org/1748-0221/12/01/C01001}{2017 JINST 12 C01001}. 2017.

\bibitem{caribou}
%\emph{CaRIBOu project webpage}, 
%\href{https://gitlab.cern.ch/Caribou/}{https://gitlab.cern.ch/Caribou/}
A. Fiergolski, \emph{A multi-chip data acquisition system based on a heterogeneous system-on-chip platform},
\href{https://cds.cern.ch/record/2272077/files/CLICdp-Conf-2017-012.pdf}{CLICdp-Conf-2017-012}. 2017.

\vspace{-0.5mm}

\bibitem{MunkerThesis}
M. Munker, \emph{Test beam and simulation studies on High Resistivity CMOS pixel sensors},
\href{https://cds.cern.ch/record/2644054}{CERN-THESIS-2018-202}. 2018.

\end{spacing}

\end{thebibliography}
